\documentclass[10pt,a4paper]{article}
\usepackage{graphicx}
\usepackage{subfigure}
\usepackage{amsmath,amsthm}
\usepackage{amssymb,amsfonts}
\usepackage{threeparttable}
\usepackage{multirow}
\usepackage{booktabs}
\usepackage{authblk}
\usepackage{url}

\newtheorem{definition}{\textbf{Definition}}

\title{Comparison on gait characteristics between controlled and free-living conditions in old adults}
\author{Jian MA\thanks{Email: majian@hitachi.cn}}
\affil{Hitachi China Research Laboratory}

\begin{document}

\maketitle

\begin{abstract}
	\noindent
	Gait is an important biomarker of functional conditions and gait characteristics can help us assessing health conditions and managing progression of diseases. Most of the existing research study the gait in controlled condition, such as clinical tests. In this paper, we study the gait characteristics in free-living conditions in old adults and compare them with that in controlled conditions, i.e., Timed Up and Go (TUG) test. 65 subjects (12 patients with mobility impairment
and 53 healthy controls) are recruited from elderly nursing institutions. The video data are collected from them in TUG test and free-living conditions and the 9 gait characteristics, including gait speed, are extracted from the data. Two-sample tests and independence test based on copula entropy are conducted on the extracted data to compare the characteristics in two conditions. Comparison results show that gait characteristics, such as gait speed, pace, speed variability, etc., in daily life are different from that of in TUG test. In daily life, people tend to have slow gait speed, smaller pace and speed variability, more frequent stride, and smaller acceleration range than in TUG test. We also found that gait speed, pace, and speed variability have stronger dependence with TUG score in the 3 conditions (TUG, daily life, and both) and that other 5 characteristics have stronger dependence with TUG score in both condition than in each condition. The comparison in this study suggests that TUG and daily life conditions are complementary with each other, and that TUG test can be considered as intervention on the movement state of human.
\end{abstract}

\section{Introduction}
Gait is widely considered as a vital sign of human health conditions. It reflects the functional ability of people and may vary due to different populations, diseases, or contextual interventions. Quantitative measurement of gait can help us understanding the symptom of diseases, assessing health conditions and managing progression of disorders \cite{Snijders2007,Baker2018}.

Gait characteristics are the quantitative measures of gait patterns from different aspects. From basic measurement of body movement, one can define many characteristics of gait to reflect different aspects of movement and functionality. The most common characteristics is gait speed, which is widely studied as a biomarker of movement disorders. Additionally, more spatial-temporal characteristics, such as pace, speed variables, stride time/length, can be defined. Characteristics of gait dynamics is also proposed to measure the dynamical properties of movement.

Gait are supposed to be different in controlled conditions and free-living condition \cite{Jette2006}. In controlled condition, such as clinical Timed Up and Go (TUG) test, individuals are asked to perform a designed task at their best possible functional ability and then his/her functional capacity is qualified as test scores. In free-living condition, individuals are doing daily activity freely and the performance is qualified from gait circles to reflect the capability in daily contextual environment. 

There have been many research on gait-based assessments in controlled clinical conditions. Meanwhile, assessment in free-living conditions is much desired for continuous home-based health monitoring, but the gait in daily life is different from and much complex than that in lab because there are many latent confounders that can affect gait pattern in these uncontrolled environment. Since gait are different in these two conditions, gait-based assessment derived from controlled conditions, such as fall risk assessment \cite{ma2020predicting}, cannot be applied directly to free-living conditions. To develop a gait-based assessment in free-living conditions, one need to study the difference between gait characteristics in two conditions. The following questions are to be answered: what characteristics are different? And how different? What factors cause the differences? Which characteristics and factors can be used in developed the models for gait-based assessment?

In this work, we will investigate the difference of gait characteristics between TUG test and free-living condition on old adults for developing free-living fall risk assessment. Particularly, we will test whether the gait characteristics in two conditions are different statistically and compare the gait characteristics of patients and control populations in each condition. We will also measure the dependence between gait characteristics and TUG score in the 3 conditions (TUG test, daily living, and both) with Copula Entropy (CE) to see which characteristics are mostly related with functional ability / fall risk score in different conditions.

\section{Related Work}
There are only a few research comparing gait characteristics in clinical and free-living conditions, or healthy and patients populations, as listed in Table \ref{tab:relatedwork}. 

Gait speed is the most studied characteristics in these works. Toosizadeh et al \cite{Toosizadeh2015} studied the difference of gait speed and other movement measures of Parkinson’ disease and healthy people in home and in clinic and found the objective measures mostly correlated with disease stage. In a longitudinal study, Rojer et al \cite{Rojer2021} investigated the interrelation between in-lab and in-daily gait speed of elderly people. They found that gait speed is distinct in different conditions and suggested to combine them together in predicting disease conditions. Corr\`a et al \cite{Corra2021} studied how to use gait speed in daily life to discriminate disease ON and OFF states. They found that gait speed in daily life reflects different aspects of mobility of PD and can complement that in clinic.

To discriminate patients with Parkinson’s disease and healthy people, Del Lin et al \cite{DelDin2016} investigated how different conditions and ambulatory bout length impact gait characteristics. In their study, 14 gait characteristics in 5 domains are included. Shah et al \cite{Shah2020a} compared 13 gait characteristics between Parkinson’s disease and controls, and multiple sclerosis and Control in lab gait test and daily life. They found that gait characteristics that best discriminate people with parkinson's disease and multiple sclerosis from healthy people are different between lab and daily life. Particularly, the toe-off angle and gait speed best discriminate multiple sclerosis and controls in lab and in daily life respectively, while the lumbbar coronal range of motion and foot-strike angle best discriminate Parkinson's disease and controls in lab and in daily life respectively. 

There are also work that compare in-lab usual walking/dual-task walking with daily ambulation in different populations. Shema-Shiratzky et al \cite{Shema-Shiratzky2020} investigated the gait change during community ambulation in multiple sclerosis. They found that during community ambulation, people in multiple sclerosis took fewer steps and slower sleep, with greater asymmetry, and larger stride variability than controls did, and that gait speed is significantly lower than in lab walking and similar to in lab dual-task walking. Hillel et al \cite{Hillel2019} did a similar research on elderly fallers. They concluded that gait measurements during daily life were worse than that during in lab and that daily life gait of elderly cannot be estimated from walking in lab.

In summary, one can learn that 1) gait speed is the mostly studied characteristics in the previous research and most works report that gait speed is slower in daily life than that in lab, and some tried to find the interrelation between gait speed in these two conditions for monitoring diseases; 2) all the previous work were based on measurements with wearable sensor; 3) most work studied walking or dual-task walking task in lab and in daily life, and only one considered TUG test in their research.

\begin{table}[htbp]
\centering
\caption{Comparison on the related work.}
\begin{threeparttable}
\begin{tabular}{c|c|c|c|c|c}
\toprule
Work&Population&\#Char&Conditions&Sensor&Task\\
\midrule
\cite{Toosizadeh2015}&PD/Ctrl&13&home/clinic&wearable&TUG/DA\\
\cite{Rojer2021}&Elderly&GS&home/lab&wearable&UW/DA\\
\cite{Corra2021}&PD&GS&home/lab&wearable&UW/DA\\
\cite{DelDin2016}&PD/Ctrl&14&home/clinic&wearable&UW/DA\\
\cite{Shah2020a}&PD,MS/Ctrl&13&home/lab&wearable&ISAW/DA\\
\cite{Shema-Shiratzky2020}&MS/Ctrl&5&home/lab&wearable&UW,DTW/DA\\
\cite{Hillel2019}&Old Faller&5&home/lab&wearable&UW,DTW/DA\\
Ours&Elderly&9&home/lab&camera&TUG/DA\\
\bottomrule
\end{tabular}
\begin{tablenotes}
\item[*] Abbreviations. PD: Parkinson’s Disease, MS: Multiple Sclerosis, Ctrl: Control, GS: Gait Speed, UW: Usual Walk, DTW: Dual-task Walk, ISAW: Instrumented Stand and Walk, DA: Daily Activity.
\end{tablenotes}
\end{threeparttable}
\label{tab:relatedwork}
\end{table}

\section{Methods}
\subsection{Two-sample tests}
Two-sample tests are a group of methods for testing whether the statistical properties of two independent samples are equal or not. The properties concerned are usually means or variance of two samples, or distributions of two samples. 

The T-test is a statistical method of such test for comparing the means of two samples. It is a parametric one which assumes normality and heterogeneity of variance within each sample. The Wilcoxon Rank Sum test, or Mann-Whitney (M-W) test \cite{Hollander1973}, is a non-parametric method for two-sample test for means. It is based on rank statistic and therefore without normality assumption.

The Kolmogorov-Smirnov (K-S) test \cite{Conover1971} is a non-parametric hypothesis test for determining whether two samples are come from a same distribution. It has two versions: the one-sample version compares a sample with a theoretical distribution and the two-sample version compares the cumulative distributions of two independence samples. 

\subsection{Copula Entropy}
Copula theory is a probabilistic theory on representation of multivariate dependence \cite{nelsen2007,joe2014}. According to Sklar’s theorem \cite{sklar1959}, any multivariate density function can be represented as a product of its marginals and copula density function which represents dependence structure among random variables. Please refer to \cite{ma2008} for notations.

With copula density function, Ma and Sun \cite{ma2008} defined a new mathematical concept, called Copula Entropy (CE), as follows:
\begin{definition}[Copula Entropy]
Let $\mathbf{X}$ be random variables with marginals $\mathbf{u}$ and copula density function $c$ . The CE of $\mathbf{x}$ is defined as
\begin{equation}
	H_c(\mathbf{x})=-\int_{u}{c(\mathbf{u})\log c(\mathbf{u})d\mathbf{u}}.
\end{equation}
\end{definition}
They proved that CE is equivalent to Mutual Information in information theory \cite{infobook}. CE has several ideal properties, such as multivariate, symmetric, invariant to monotonic transformation, non-positive (0 iff independent), and equivalent to correlation coefficient in Gaussian cases. It is a perfect measure for statistical independence. 

Ma and Sun \cite{ma2008} also proposed a non-parametric method for estimating CE, which composes of two simple steps: 1) estimating empirical copula density function; and 2) estimation CE from the estimated empirical copula density function. In the first step, rank statistic is used to derive empirical copula density function; in the second step, the famous KSG method for estimating entropy \cite{kraskov2004} is suggested. The proposed estimation method is rank-based and essentially to estimate the entropy of rank statistic. 

In a word, CE provides a ideal tool for testing statistical independence with a non-parametric estimation method.

\section{Data}
\subsection{Data Collection}
The data in this study were collected on 65 subjects (12 patients with mobility impairment and 53 healthy controls) recruited from elderly nursing institutions in Tianjin and Chengdu, China, whose age range at 45$\sim$84. All the participants signed informed consent. The subjects at Tianjin and part (8) of subjects at Chengdu were administrated to perform TUG tests twice a day for several times in one month and 149 tests were performed totally. For each test, a video about 2$\sim$4 minutes was recorded. Additionally, a group (22) of subjects at Chengdu are requested to record a batch of video of their domestic activities which makes 22 videos totally. All the subjects of daily video were also asked to do a TUG test to derive a TUG score accordingly. The details of the subjects are listed in Table \ref{tab:subjects}. 

\begin{table}
\centering
\caption{Subjects at Tianjin and Chengdu.}
\begin{tabular}{c|c|c|c|c}
\toprule
PoC&\multicolumn{2}{|c|}{\#subjects/patients}&\#TUG Tests&\#Daily Activity\\
\midrule
Tianjin&\multicolumn{2}{|c|}{40/4}&137&/\\
\hline
\multirow{2}{*}{Chengdu}&\multirow{2}{*}{25/8}&8/3&12&/\\
\cline{3-5}
&&22/6&/&22\\
\midrule
Total&\multicolumn{2}{|c|}{65/12}&149&22\\
\bottomrule
\end{tabular}
\label{tab:subjects}
\end{table}

\begin{figure}
	\centering
	\includegraphics[width =0.9\textwidth]{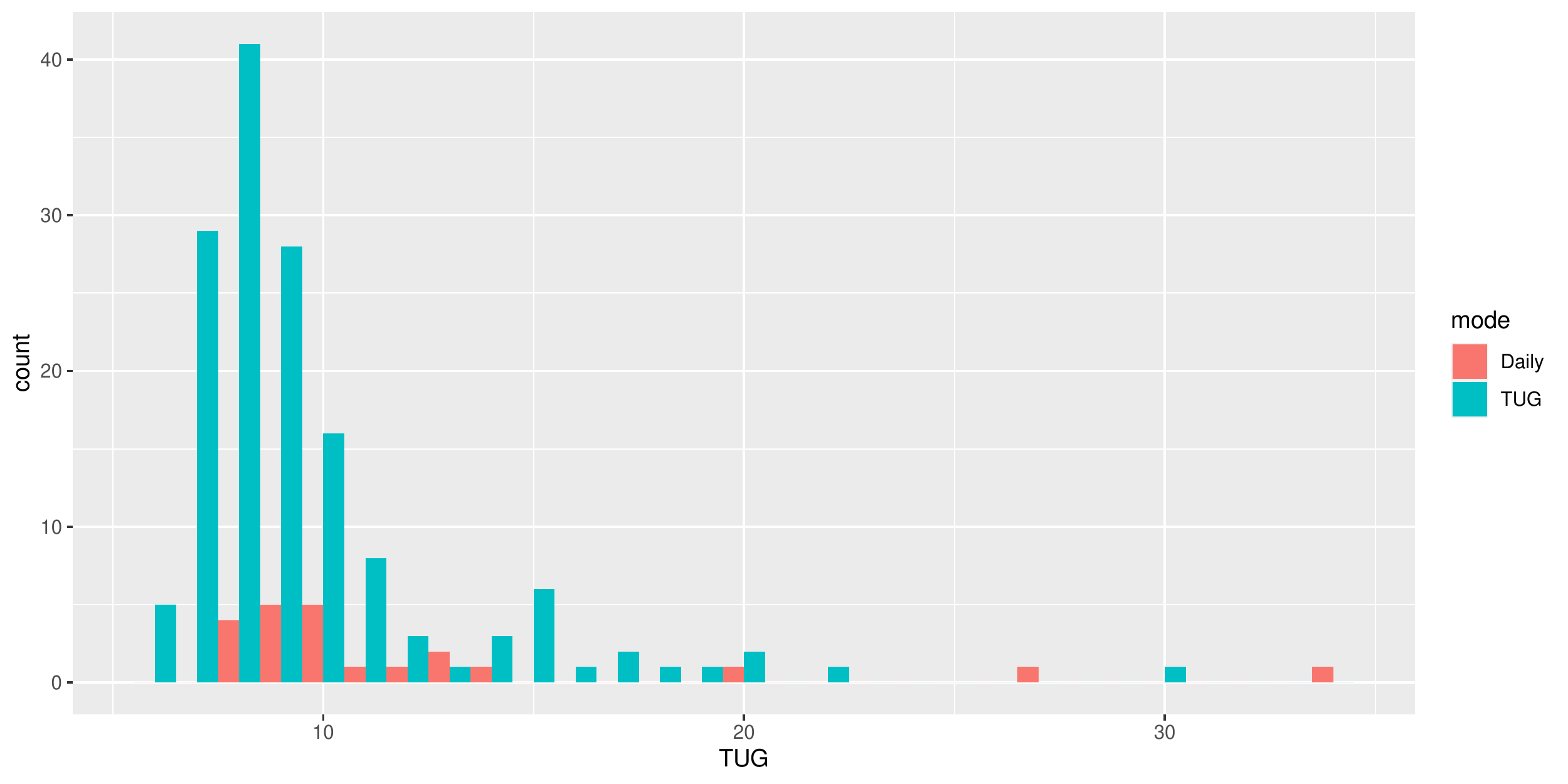}
	\caption{Histogram of the TUG scores of the two modes (TUG test and daily living).}
	\label{fig:tughist}
\end{figure}

The distribution of TUG score of two conditions are shown in Figure \ref{fig:tughist}. It can be learned from the figure that most of the subjects are healthy people and that the distribution of the TUG scores of two conditions are similar.

\subsection{Data Preprocessing}
The video data collected above were first edited to preserve the interested contents: the video of TUG tests and the video of free walking (speed$>$0) in free-living scenario. From the edited video data of TUG test and daily activities, 9 gait characteristics (See Table \ref{tab:chardef} for definitions) were extracted with the method proposed in Ref \cite{Li2019}  originally. Then a sample set were derived from the original data of gait characteristics, of which each sample composed of 9 characteristics generated on ambulatory bouts with length set as 15s. The process that generates these samples are controlled with two parameters: bout length (=15s) and the step between two neighbour bouts (=3s). The samples such generated can represent the gait characteristics of the subjects on a proper time scale better than the original characteristics that generated on a much shorter ambulatory bout. And then each sample is attached with the TUG score of the subject of the corresponding video. Finally, we obtained 607 samples totally (472 samples from the videos of TUG test and 135 samples from that of daily activities).

\begin{table}
\centering
\caption{Definitions of the extracted gait characteristics from video data.}
\begin{tabular}{l|l}
\toprule
Characteristics&Definition\\
\midrule
Gait velocity (Speed)&Speed of body movement\\
\hline
Speed variability&\parbox{0.6\columnwidth}{The standard deviation of the stride speeds with exclusion of the highest and lowest 10\%}\\
\hline
Stride time&\parbox{0.6\columnwidth}{The time between one peak and the second-next peak}\\
\hline
Stride time variability&\parbox{0.6\columnwidth}{The standard deviation of stride times}\\
\hline
Stride frequency&\parbox{0.6\columnwidth}{The median of the modal frequency for the ML and half the modal frequencies for the V and AP directions}\\
\hline
Movement intensity&\parbox{0.6\columnwidth}{The standard deviation of the acceleration signals}\\
\hline
Low-frequency percentage&\parbox{0.6\columnwidth}{Summed power up to a threshold frequency divided by total power}\\
\hline
Acceleration range&\parbox{0.6\columnwidth}{Difference between the minimum and maximum acceleration}\\
\hline
Step length (Pace)&Length of one step\\
\bottomrule
\end{tabular}
\label{tab:chardef}
\end{table}

\section{Experiments}
In this research, we did 4 group of experiments. To compare gait characteristics between TUG test and daily life, we first did M-W test on them to test whether the means of gait characteristics in two conditions are different. Since mean can only reflect limited information of gait characteristics, we then test whether the distributions of 9 gait characteristics in these two conditions are different with the non-parametric K-S test. 

We also conducted experiments on studying the interrelation between gait characteristics in 3 conditions (TUG test, daily life, and both). The relationship is measured with correlation coefficient. For each condition, we studied the interrelation between gait characteristics in 3 (sub-)populations (patients, controls and the whole).

To develop gait-based assessment models, we also investigated the relationship between gait characteristics and TUG score in 3 conditions to see how these relations change with different conditions. The dependence between characteristics and TUG scores will be measured with CE.

\section{Results}
\begin{figure}
	\centering
	\subfigure[Speed]{\includegraphics[width=0.45\textwidth]{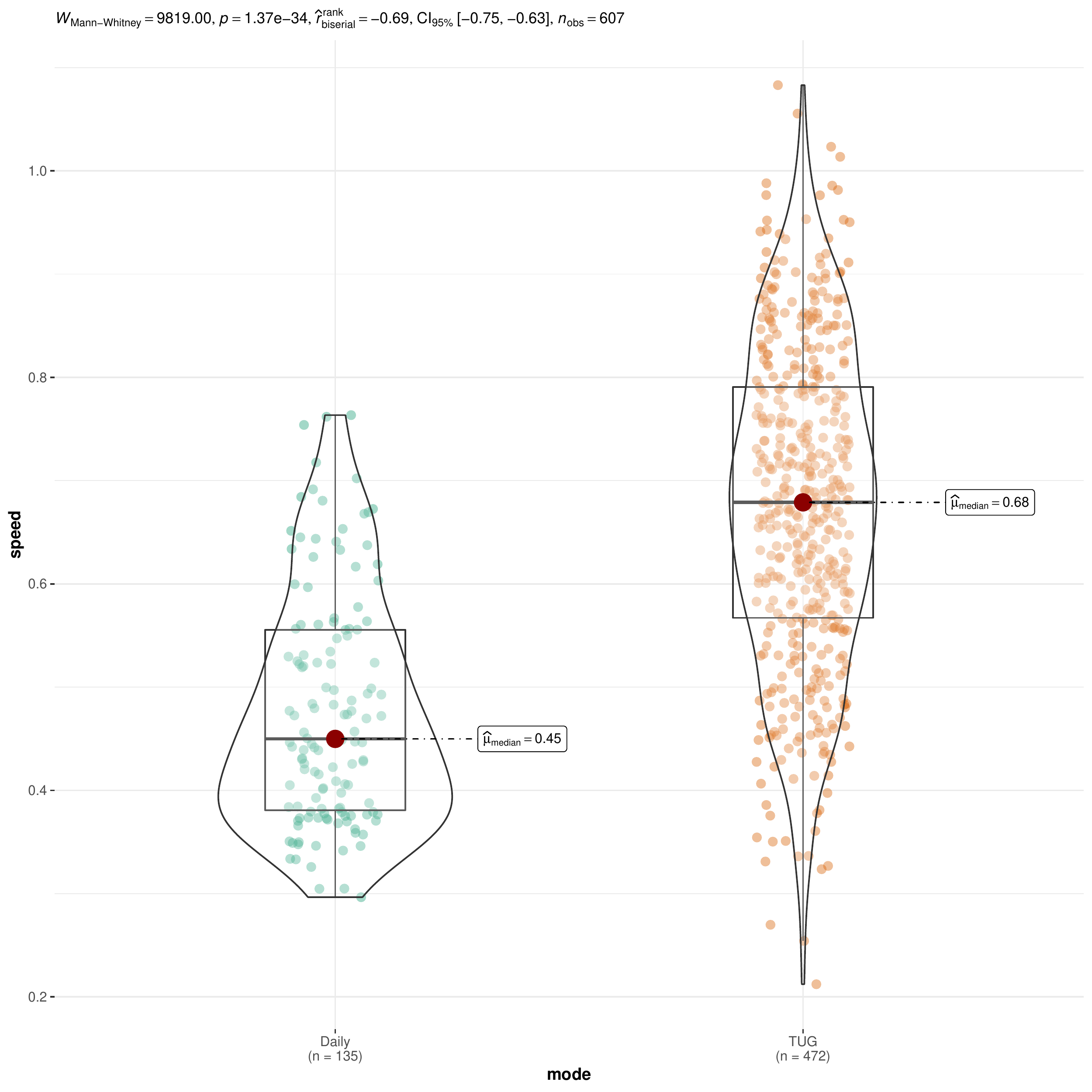}}
	\subfigure[Pace]{\includegraphics[width=0.45\textwidth]{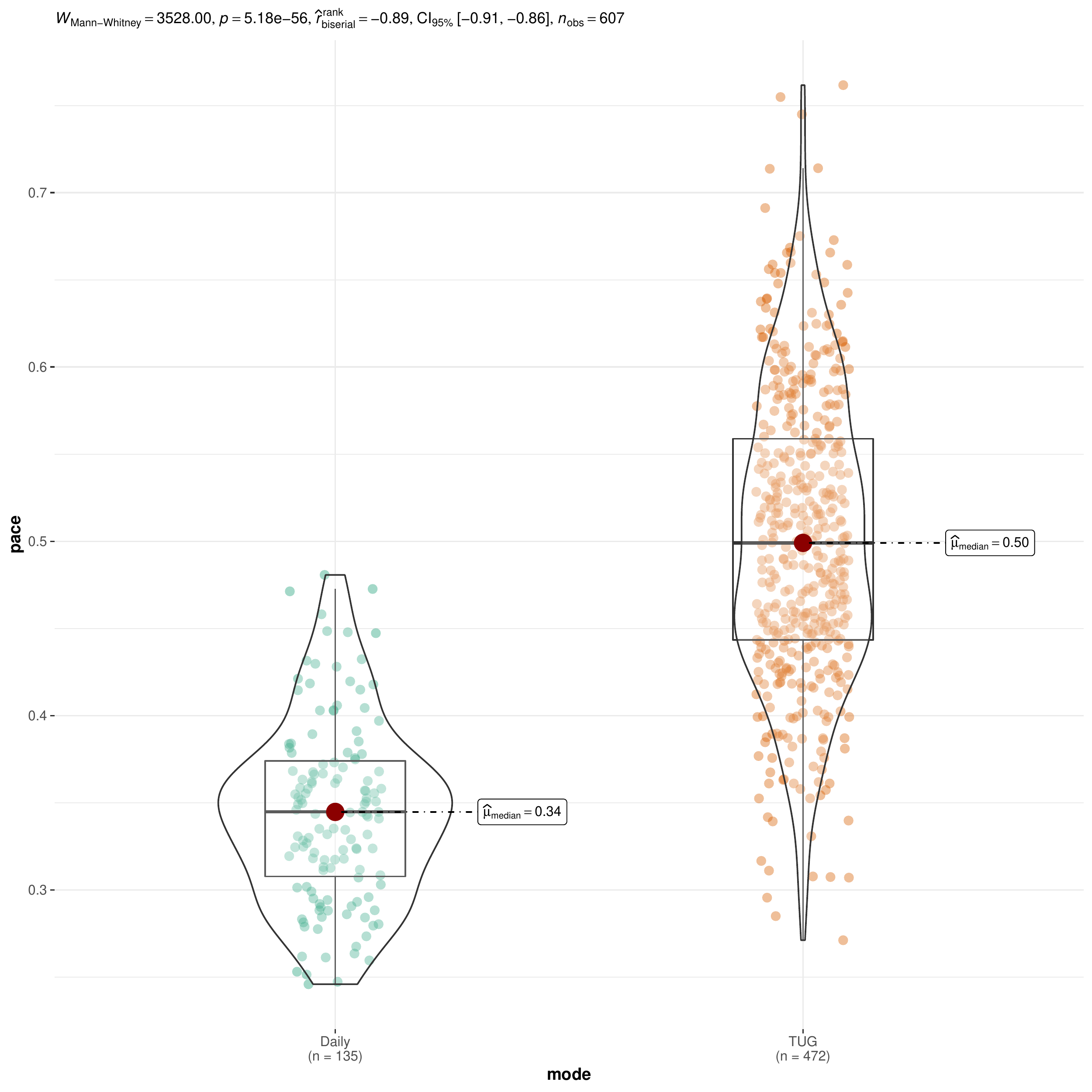}}
	\subfigure[Speed Variability]{\includegraphics[width=0.45\textwidth]{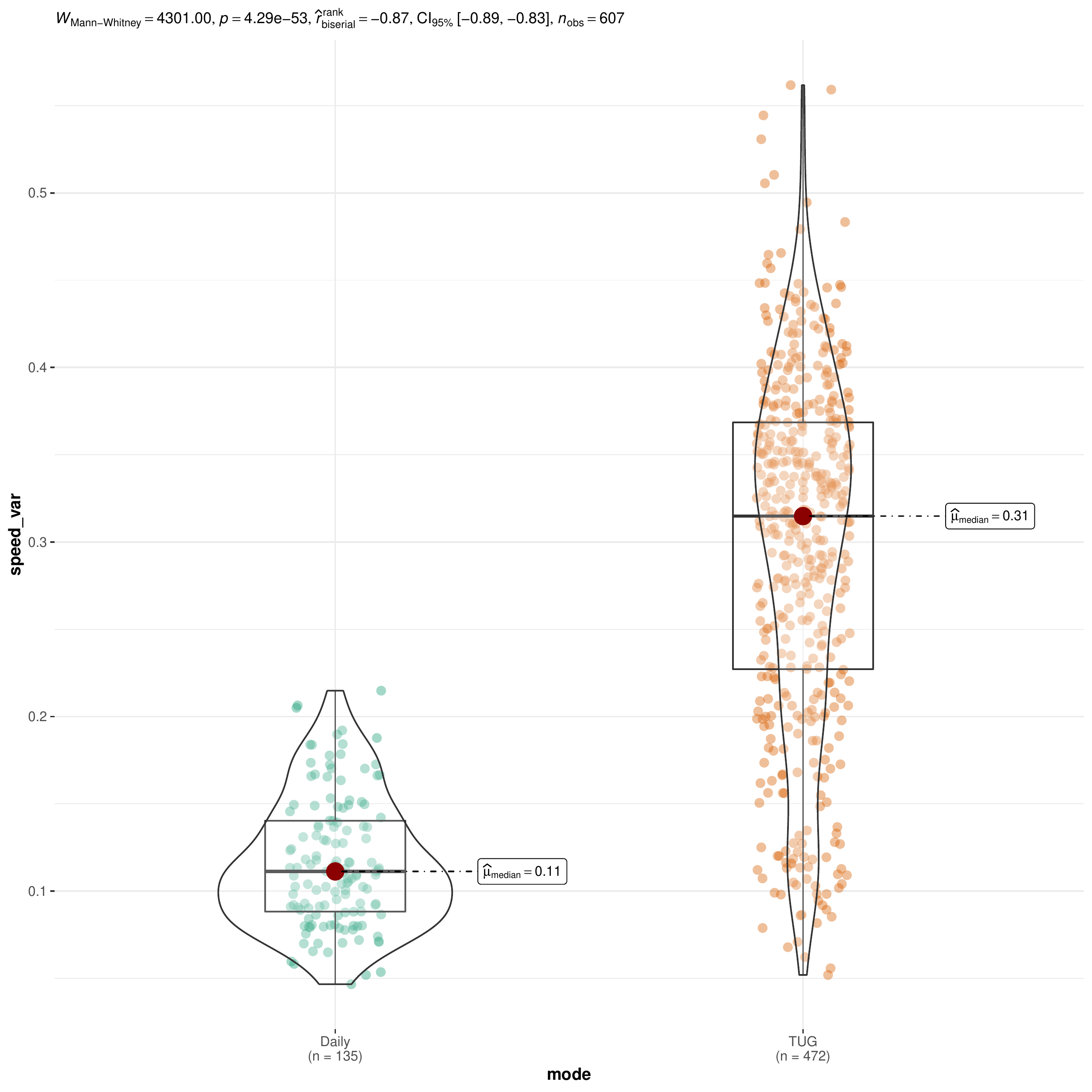}}
	\subfigure[Stride time variability]{\includegraphics[width=0.45\textwidth]{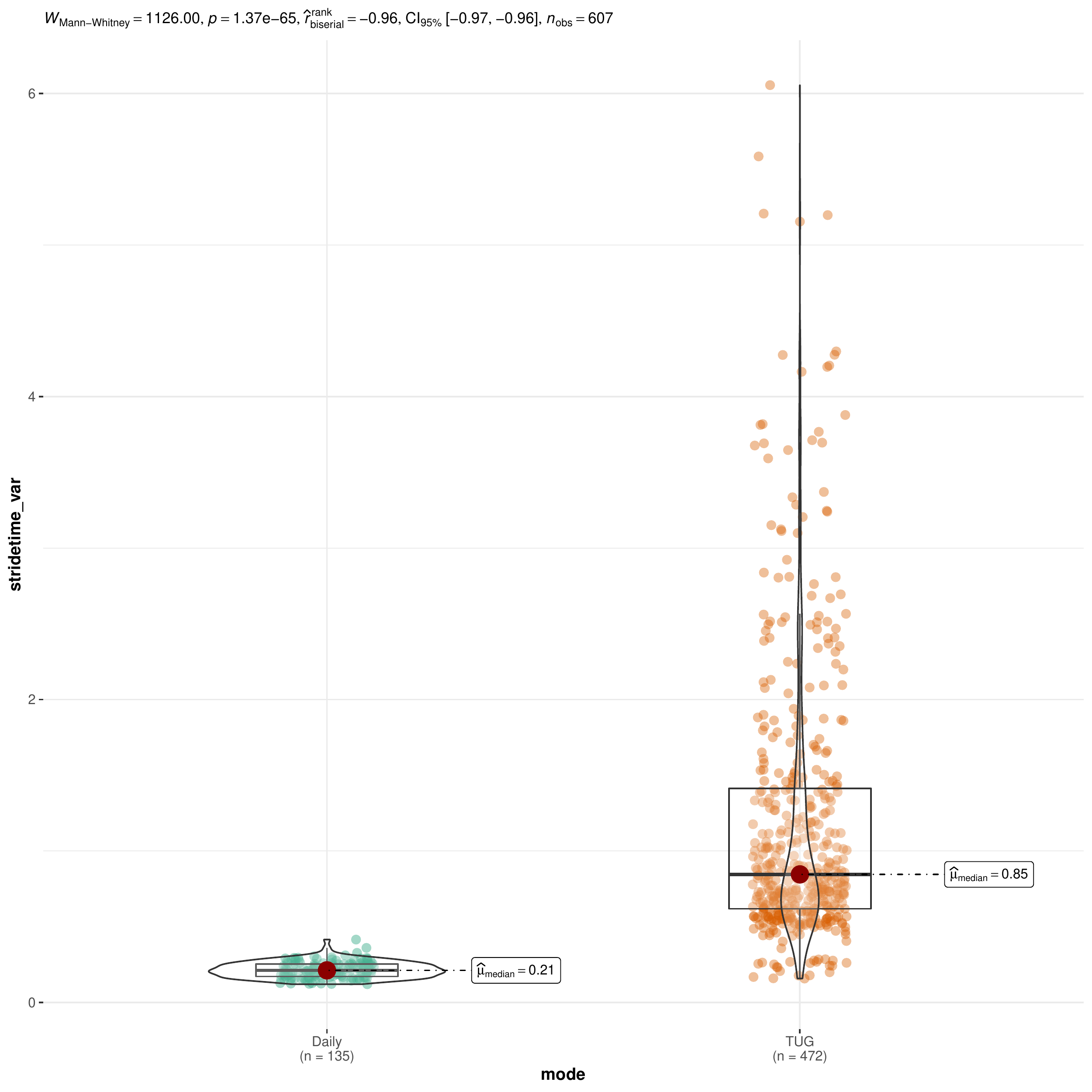}}
	\caption{Results of Mann-Whitney test on the 4 gait characteristics.}
	\label{fig:twosample}
\end{figure}

\begin{figure}
	\centering
	\includegraphics[width=\textwidth]{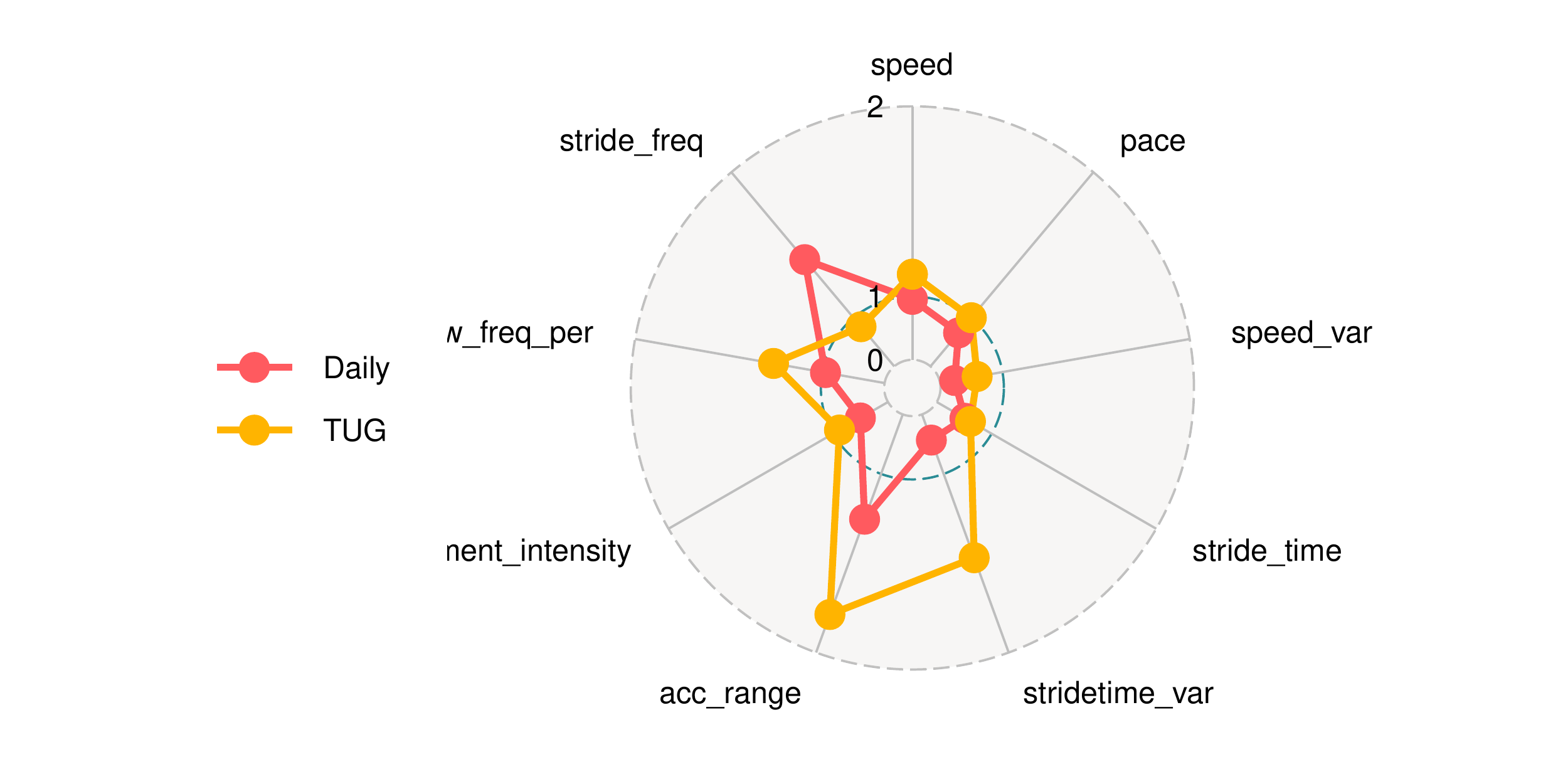}
	\caption{Comparison of the means of 9 gait characteristics of two conditions.}
	\label{fig:radar}
\end{figure}

The results of two-sample tests are illustrated in Figure \ref{fig:twosample} and listed in Table \ref{tab:twosample}. The mean and variance of 9 gait characteristics in two conditions are first calculated and compared, as listed in Table \ref{tab:twosample} and illustrated in Figure \ref{fig:radar}. It can be learned from them that gait speed, pace, and speed variability in daily life are lower than that of in TUG test. This confirmed the results of the previous works \cite{Shah2020a}. Besides, other characteristics share the similar results except stride frequency which is larger in daily life than in TUG test. This means that the statistical properties of the 9 gait characteristics changes as movement condition changes.

\begin{table}
	\centering
	\caption{Results of two-sample tests on the gait characteristics of two conditions.}
	\begin{tabular}{l|c|c|c|c|c|c}
		\toprule
		\multirow{2}{*}{Characteristics}&\multicolumn{2}{|c|}{Mean and Var}&\multicolumn{2}{|c|}{K-S test}&\multicolumn{2}{|c}{M-W test}\\
		\cline{2-7}
		&TUG&Daily&Statistic&P-value&Statistic&P-value\\
		\midrule
		Speed&0.68$\pm$0.02&0.48$\pm$0.01&0.559&0&53901&1.37e-14\\
		Pace&0.50$\pm$0.01&0.34$\pm$0.00&0.760&0&60192&5.18e-56\\
		Speed var.&0.30$\pm$0.01&0.17$\pm$0.00&0.810&0&59419&4.29e-53\\
		Stride time&0.31$\pm$2.10&0.26$\pm$0.00&0.689&0&51583&4.96e-28\\
		Stride time var.&1.20$\pm$0.90&0.22$\pm$0.00&0.947&0&62594&1.37e-65\\
		Acceleration range&1.68$\pm$1.54&0.88$\pm$0.11&0.437&0&44653&1.08e-12\\
		Movement intensity&0.44$\pm$0.11&0.25$\pm$0.01&0.398&6.77e-15&42811&1.10e-09\\
		Low freq. perc.&0.89$\pm$0.01&0.47$\pm$0.01&0.951&0&61849.5&1.54e-62\\
		Stride freq.&0.41$\pm$0.08&1.10$\pm$0.06&0.953&0&2057&8.66e-62\\
		\bottomrule
	\end{tabular}
\label{tab:twosample}
\end{table}

The M-W test of the 9 gait characteristics present p-value smaller than 0.05, which also suggest that the means of 9 gait characteristics in two conditions are all different statistically. the 8 K-S tests did not present p-value due to ties in samples. However, we can still learn from the figure that the distributions of the 9 gait characteristics in two conditions are different.

\begin{figure}
	\centering
	\includegraphics[width=\textwidth]{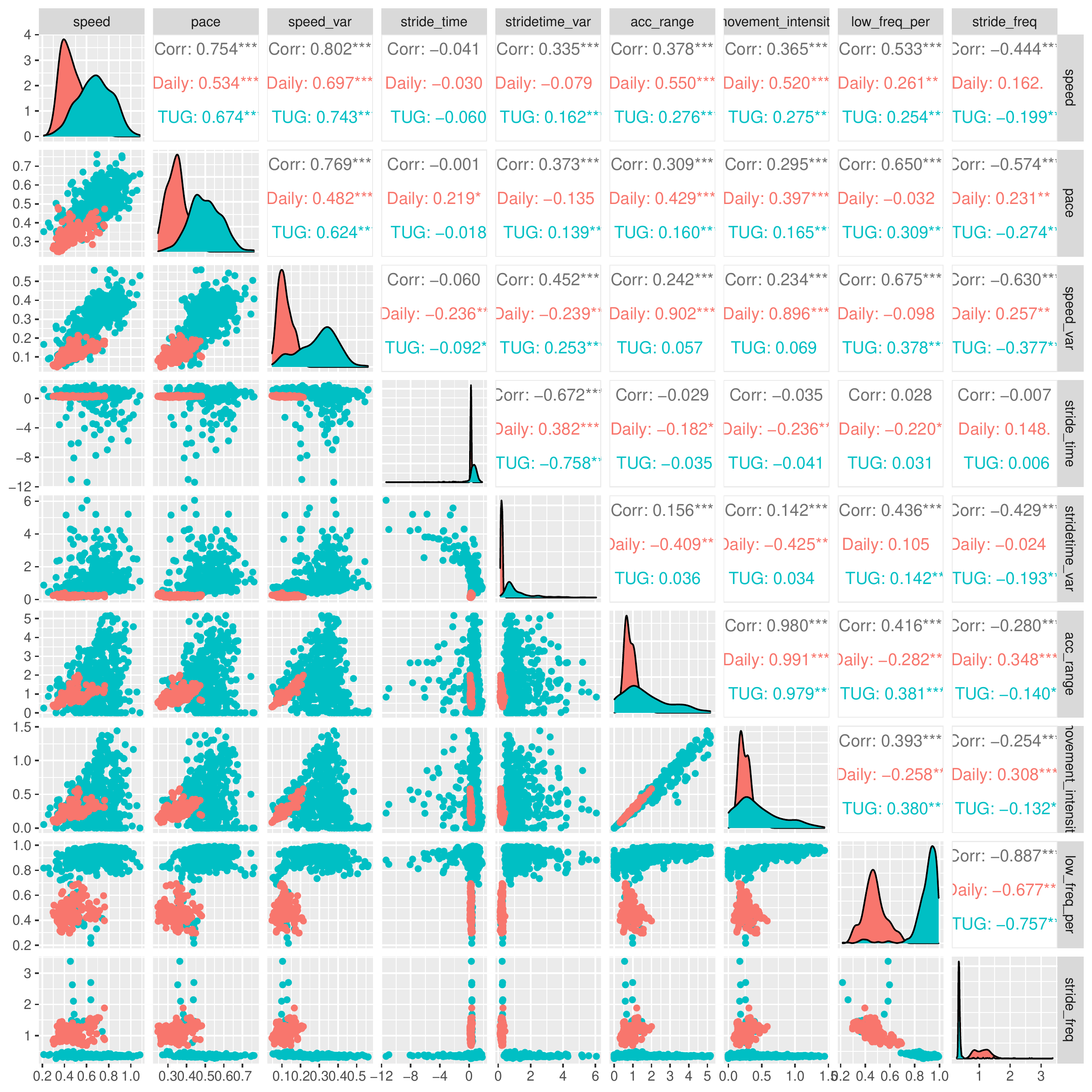}
	\caption{Joint plots of 9 gait characteristics derived from two conditions (TUG and daily life).}
	\label{fig:pairs}
\end{figure}

The joint plots of the 9 gait characteristics in two conditions are shown in Figure \ref{fig:pairs}. From it, one can learn that the distribution of gait characteristics in two conditions are distinct. We can also learn that the interrelations between gait speed, pace, and speed variability in each condition (TUG and daily life) are weaker than that in both together as the measure of correlation coefficients indicate, which means that the interrelation between these 3 characteristics are becoming stronger if considering both conditions together. We can also learn from Figure \ref{fig:pairs} that the other 6 characteristics have 3 interrelated groups by correlation coefficients: stride time and stride time variability, acceleration range and movement intensity, and low frequency percentage and stride frequency.

\begin{figure}
	\centering
	\includegraphics[width=\textwidth]{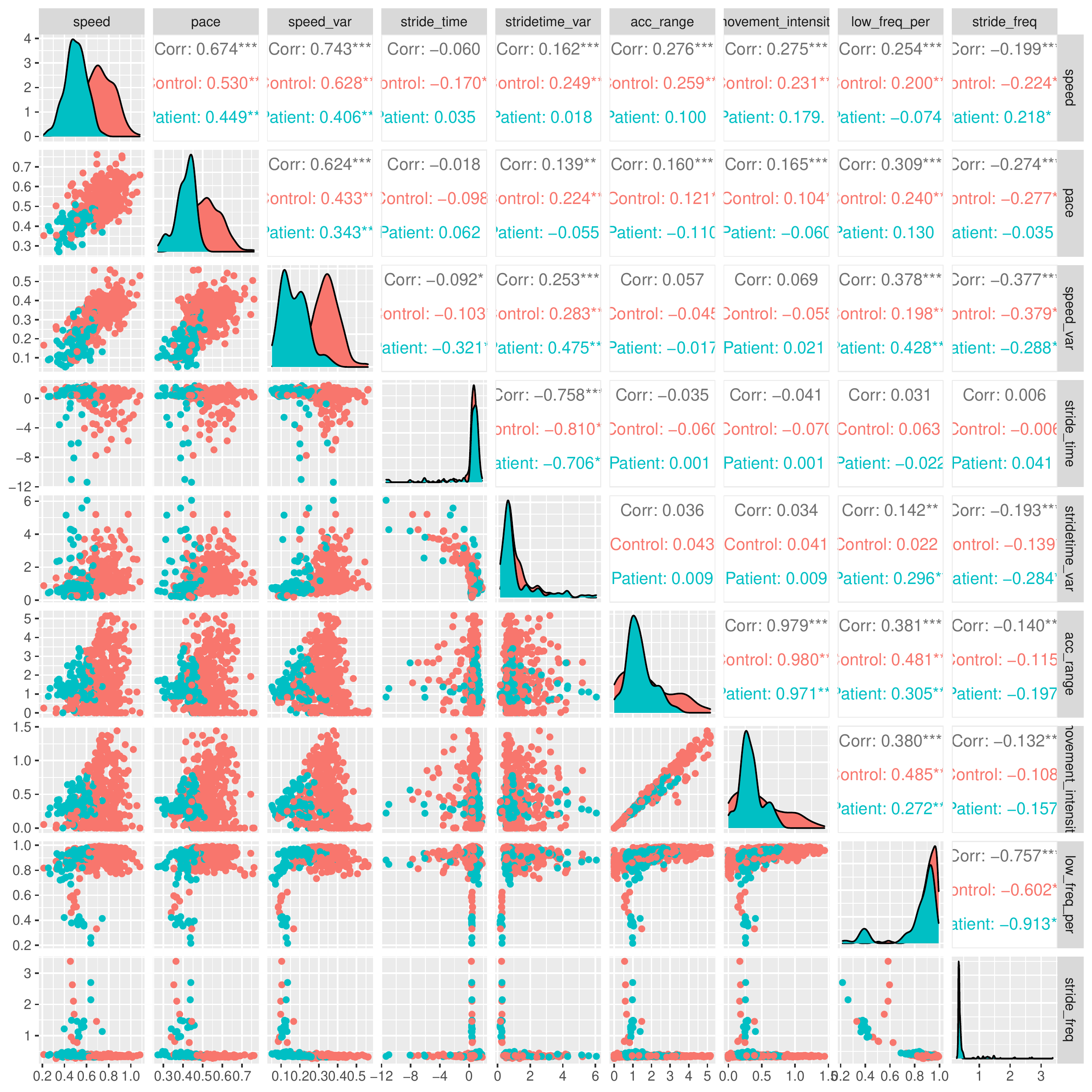}
	\caption{Joint plot of 9 gait characteristics of two sub-populations (patients and controls) in TUG test.}
	\label{fig:pairs1tug}
\end{figure}

\begin{figure}
	\centering
	\includegraphics[width=\textwidth]{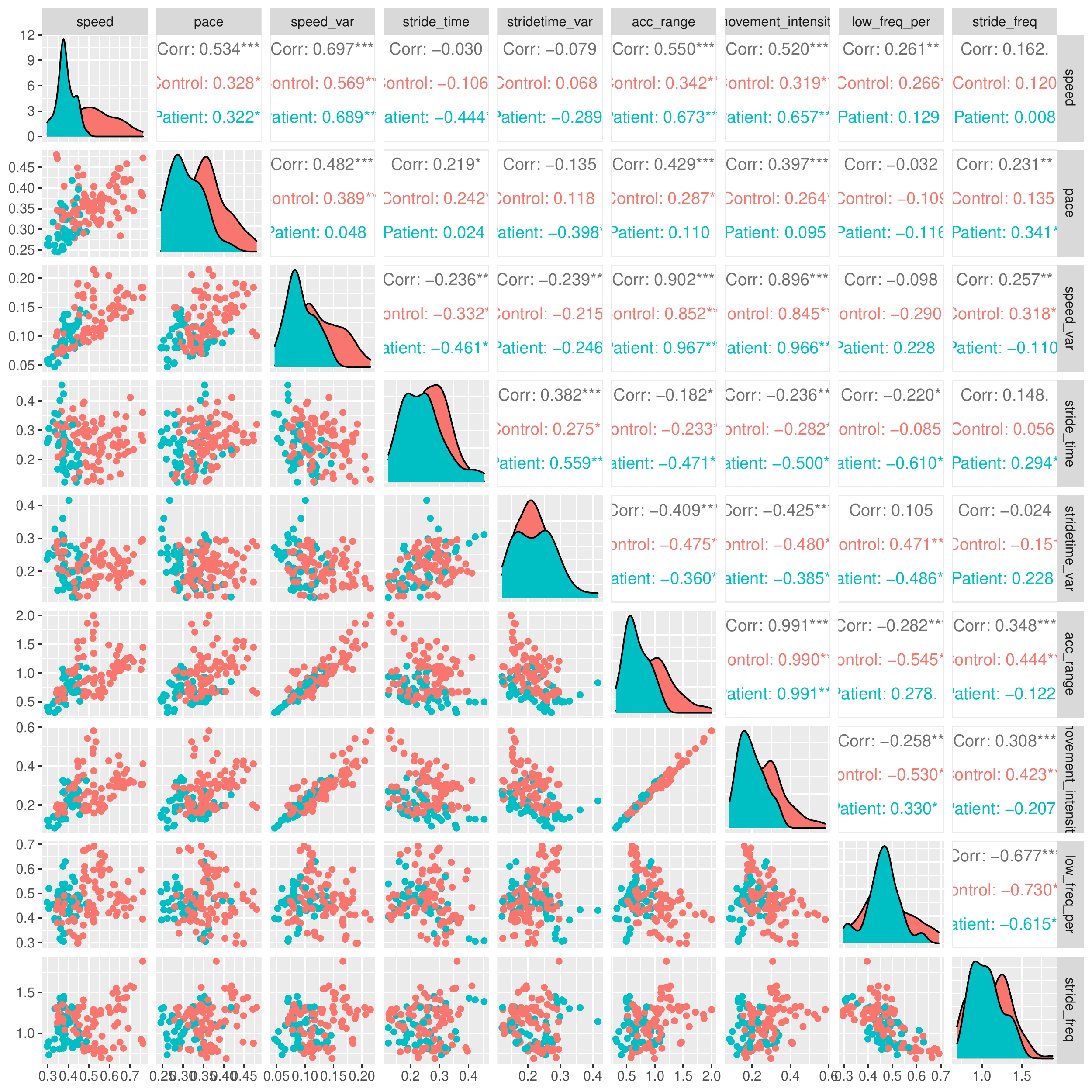}
	\caption{Joint plot of 9 gait characteristics of two sub-populations (patients and controls) in free-living condition.}
	\label{fig:pairs1daily}
\end{figure}

The joint plots of the 9 gait characteristics in TUG test and daily life conditions are shown in Figure \ref{fig:pairs1tug} and Figure \ref{fig:pairs1daily} respectively. In each figure, the distribution of gait characteristics of patients and controls sub-populations are illustrated separately and the interrelation are measured with correlation coefficients. We can learn from Figure \ref{fig:pairs1tug} and \ref{fig:pairs1daily} that gait speed, pace, and speed variability of patients and control sub-populations are different in terms of distribution and that the interrelation between these 3 characteristics measured by correlation coefficients become stronger in the whole population than that in each sub-population.

\begin{figure}
	\centering
	\includegraphics[width=0.95\textwidth]{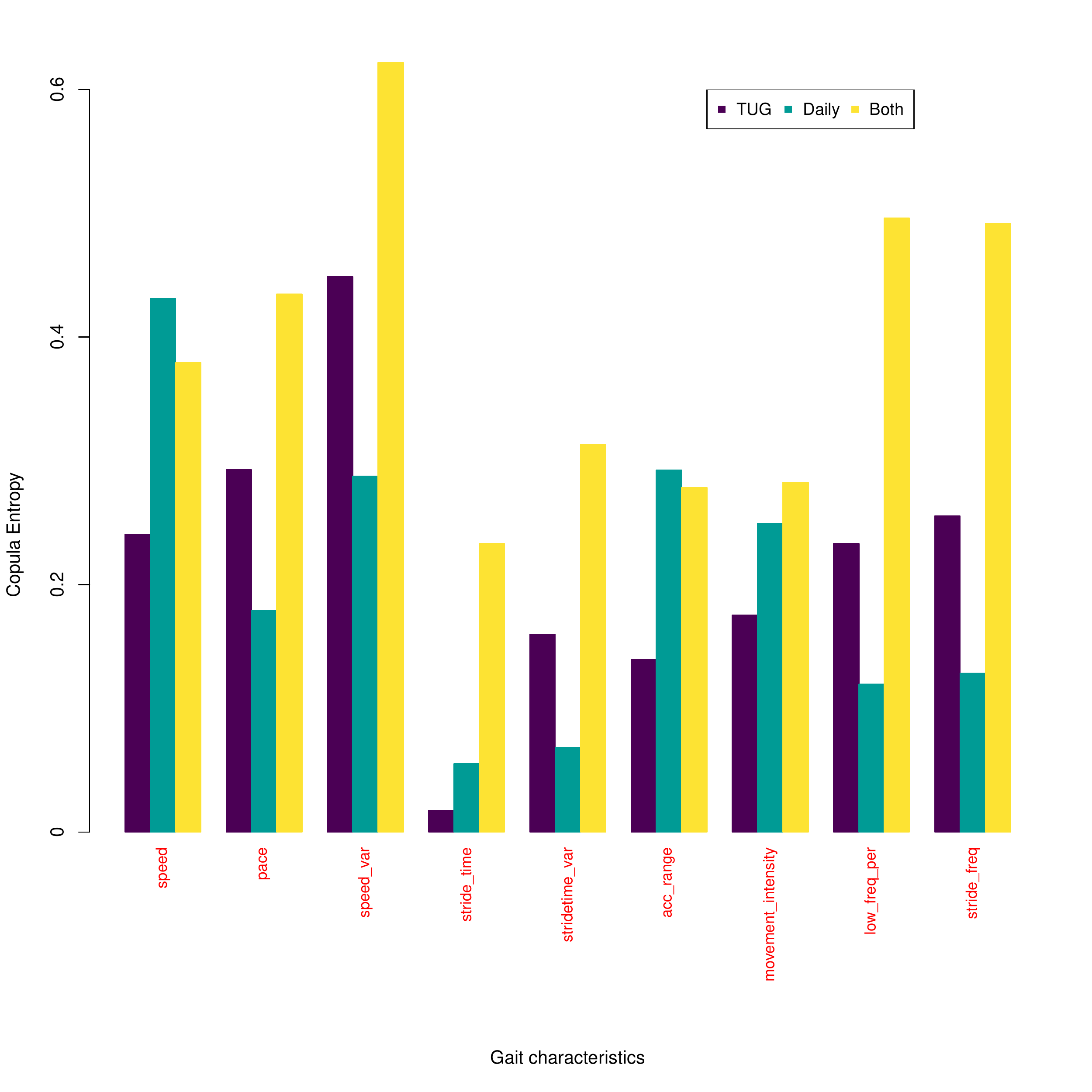}
	\caption{Dependence between gait characteristics and TUG score in 3 conditions measured by CE.}
	\label{fig:dmi}
\end{figure}

The dependence between gait characteristics and TUG score in 3 conditions (TUG test, daily life, and both) are illustrated in Figure \ref{fig:dmi}. It can be learned from it that gait speed, pace, and speed variability are the 3 characteristics that have strong dependence with TUG score in TUG and the both conditions while gait speed and speed variability also have strong dependence with TUG score in daily life condition. There are 5 gait characteristics, including speed variability, stride time, stride time variability, low frequency percentage, and stride frequency, that has much stronger dependence with TUG score in both conditions than that in each condition alone.

\section{Discussion}
In this research, bout length and interval between bouts are two hyper-parameters in the process of generating gait characteristics. The interval between bouts controls the sample size and bout length controls the quality of gait characteristics. Bout length was set as 15s in this work after hyper-parameter tuning process. With this value of bout length, we can generate gait characteristics that can reflect the characteristics of movement and functionality of different populations in different conditions at optimal level while has a large enough sample for analysis. This value is also a good choice according to the previous research \cite{DelDin2016}, in which bout length is suggested to be larger than 10s or longer for better between group difference of gait characteristics. Another research by Corr\`a et al \cite{Corra2021} also generates gait speed per bout on a interval longer than 15s so as to guarantee the accuracy of further analysis.

Gait characteristics in this research are derived from video data collected with 3D camera as biosensor. In this process, video data are first fed into pose estimation algorithm to generate the pose series and then the generated pose series are used to derive gait characteristics. This process is reliable since the accuracy of pose estimation algorithm have been evaluated in previous research \cite{Needham2021,Mehdizadeh2021}. To the best of our knowledge, this is the first work that compared the difference of gait characteristics between in-lab and in daily life conditions with 3D vision sensor. As contrast, the related work based on wearable sensors may have several disadvantages, such as complex algorithms, measurement error, susceptible to noise \cite{Muro-de-la-Herran2014}, and hence less reliable compared with ours.

Experimental results showed that the 9 gait characteristics in TUG test and daily life are different in terms of mean and distribution as a whole as tested with K-S test and M-W test. We can learn that gait speed, pace and speed variability are larger in TUG test than in daily life. Previously, Toosizadeh et al \cite{Toosizadeh2015} investigated the difference of 13 gait characteristics, including gait speed, stride length, and speed variability, between PD and healthy control group in iTUG and daily life conditions. But they did not compare gait speed, pace, and speed variability between TUG test and daily life conditions. Additionally, several new gait characteristics in our study were not considered in previous research, including stride time, stride time variability, acceleration range, movement intensity, low frequency percentage, and stride frequency. We found that stride frequency is larger in daily life than in TUG test condition and that acceleration range is lower in daily life than in TUG test condition, which means people tend to walk with more frequent but small step in daily life than in TUG test condition while with larger acceleration. These comparison helps us understanding the difference of gait between two conditions more than before.

With these comparisons, we can learn more on how people perform in different conditions. In this sense, TUG test can be considered as a intervention that change the movement state of both patients and healthy people. In TUG test, people intend to perform at best possible capacity of his/her body functionality while in daily life, they return to ‘normal’ state. This means that the underlying relationship between gait characteristics and functional outcomes can be modeled as function with intervention as a parameter. From Figure \ref{fig:dmi}, we can learn that the dependence between several gait characteristics and TUG score become stronger in the both condition than in each condition. It implies these characteristics are potential predictors of the outcome of functional ability, i.e., TUG score.

\section{Conclusions}
Gait is an important biomarker of functional conditions and gait characteristics that measure different aspects of gait can help us assessing health conditions and managing progression of diseases. In this paper, we study the gait characteristics in free-living conditions in old adults and compare them with that in TUG test. A group of old adults (12 patients with mobility impairment
and 53 healthy controls) are recruited to collect video data from TUG test and their daily life with 3D camera. The 9 gait characteristics, including gait speed, are extracted from the data. Two types of two-sample tests (K-S test and M-W test) are conducted to test the difference of gait characteristics between in TUG test and in daily life conditions. And independence test based on CE are conducted to compare the dependence between gait characteristics and TUG score in these two conditions. Comparison results show that gait characteristics, such as gait speed, pace, speed variability, etc., in daily life are different from that of in TUG test. In daily life, people tend to have slow gait speed, smaller pace and speed variability, more frequent stride, and smaller acceleration range than in TUG test. We also found that gait speed, pace, and speed variability have stronger dependence with TUG score in the 3 conditions (TUG, daily life, and both) and that other 5 characteristics have stronger dependence with TUG score in both conditions than in each condition. 

The comparison in this study suggests that TUG and daily life conditions provide complementary information with each other and that TUG test can be considered as intervention on the movement state of human which change the body for 'normal' state to a 'test' state. This insight will help us to develop gait-based assessment in daily life condition.

\section*{Acknowledgement}
The author thanks Lin Xiaolie for providing video data.

\bibliographystyle{unsrt}
\bibliography{diff}

\begin{thebibliography}{10}

\bibitem{Snijders2007}
Anke~H. Snijders, Bart~P. van~de Warrenburg, Nir Giladi, and Bastiaan~R. Bloem.
\newblock Neurological gait disorders in elderly people: clinical approach and
  classification.
\newblock {\em The Lancet Neurology}, 6(1):63--74, January 2007.

\bibitem{Baker2018}
Jessica~M. Baker.
\newblock Gait disorders.
\newblock {\em The American Journal of Medicine}, 131(6):602--607, June 2018.

\bibitem{Jette2006}
Alan~M Jette.
\newblock {Toward a Common Language for Function, Disability, and Health}.
\newblock {\em Physical Therapy}, 86(5):726--734, 05 2006.

\bibitem{ma2020predicting}
Jian Ma.
\newblock Predicting tug score from gait characteristics with video analysis
  and machine learning.
\newblock {\em arXiv preprint arXiv:2003.00875}, 2020.

\bibitem{Toosizadeh2015}
Nima Toosizadeh, Jane Mohler, Hong Lei, Saman Parvaneh, Scott~J. Sherman, and
  Bijan Najafi.
\newblock Motor performance assessment in parkinson’s disease: Association
  between objective in-clinic, objective in-home, and subjective/semi-objective
  measures.
\newblock {\em PLoS ONE}, 10(4):e0124763, 2015.

\bibitem{Rojer2021}
A.~G.~M. Rojer, A.~Coni, S.~Mellone, J.~M. Van~Ancum, B.~Vereijken, J.~L.
  Helbostad, K.~Taraldsen, S.~Mikolaizak, C.~Becker, K.~Aminian, M.~C.
  Trappenburg, C.~G.~M. Meskers, A.~B. Maier, and M.~Pijnappels.
\newblock Robustness of in-laboratory and daily-life gait speed measures over
  one year in high functioning 61- to 70-year-old adults.
\newblock {\em Gerontology}, 67:650--659, 2021.

\bibitem{Corra2021}
Marta~Francisca Corrà, Arash Atrsaei, Ana Sardoreira, Clint Hansen, Kamiar
  Aminian, Manuel Correia, Nuno Vila-Chã, Walter Maetzler, and Luís Maia.
\newblock Comparison of laboratory and daily-life gait speed assessment during
  on and off states in parkinson’s disease.
\newblock {\em Sensors}, 21(12), 2021.

\bibitem{DelDin2016}
Silvia Del~Din, Alan Godfrey, Brook Galna, Sue Lord, and Lynn Rochester.
\newblock Free-living gait characteristics in ageing and parkinson’s disease:
  impact of environment and ambulatory bout length.
\newblock {\em Journal of NeuroEngineering and Rehabilitation}, 13(1):46, May
  2016.

\bibitem{Shah2020a}
Vrutangkumar~V. Shah, James McNames, Martina Mancini, Patricia Carlson-Kuhta,
  Rebecca~I. Spain, John~G. Nutt, Mahmoud El-Gohary, Carolin Curtze, and Fay~B.
  Horak.
\newblock Laboratory versus daily life gait characteristics in patients with
  multiple sclerosis, parkinson’s disease, and matched controls.
\newblock {\em Journal of NeuroEngineering and Rehabilitation}, 17(1):159,
  December 2020.

\bibitem{Shema-Shiratzky2020}
Shirley Shema-Shiratzky, Inbar Hillel, Anat Mirelman, Keren Regev, Katherine~L.
  Hsieh, Arnon Karni, Hannes Devos, Jacob~J. Sosnoff, and Jeffrey~M. Hausdorff.
\newblock A wearable sensor identifies alterations in community ambulation in
  multiple sclerosis: contributors to real-world gait quality and physical
  activity.
\newblock {\em Journal of Neurology}, 267(7):1912--1921, July 2020.

\bibitem{Hillel2019}
Inbar Hillel, Eran Gazit, Alice Nieuwboer, Laura Avanzino, Lynn Rochester,
  Andrea Cereatti, Ugo~Della Croce, Marcel~Olde Rikkert, Bastiaan~R. Bloem,
  Elisa Pelosin, Silvia Del~Din, Pieter Ginis, Nir Giladi, Anat Mirelman, and
  Jeffrey~M. Hausdorff.
\newblock Is every-day walking in older adults more analogous to dual-task
  walking or to usual walking? elucidating the gaps between gait performance in
  the lab and during 24/7 monitoring.
\newblock {\em European Review of Aging and Physical Activity}, 16(1):6, May
  2019.

\bibitem{Hollander1973}
Myles Hollander and Douglas~A. Wolfe.
\newblock {\em Nonparametric Statistical Methods}.
\newblock New York: John Wiley \& Sons, 1973.

\bibitem{Conover1971}
William~J. Conover.
\newblock {\em Practical Nonparametric Statistics}.
\newblock New York: John Wiley \& Sons, 1971.

\bibitem{nelsen2007}
Roger~B Nelsen.
\newblock {\em An introduction to copulas}.
\newblock Springer Science \& Business Media, 2007.

\bibitem{joe2014}
Harry Joe.
\newblock {\em Dependence modeling with copulas}.
\newblock CRC press, 2014.

\bibitem{sklar1959}
Abe Sklar.
\newblock Fonctions de repartition an dimensions et leurs marges.
\newblock {\em Publications de l'Institut de statistique de l'Universit\'e de
  Paris}, 8:229--231, 1959.

\bibitem{ma2008}
Jian Ma and Zengqi Sun.
\newblock Mutual information is copula entropy.
\newblock {\em Tsinghua Science \& Technology}, 16(1):51--54, 2011.
\newblock See also arXiv preprint arXiv:0808.0845 (2008).

\bibitem{infobook}
Thomas~M Cover and Joy~A Thomas.
\newblock {\em Elements of information theory}.
\newblock John Wiley \& Sons, 2012.

\bibitem{kraskov2004}
Alexander Kraskov, Harald St{\"o}gbauer, and Peter Grassberger.
\newblock Estimating mutual information.
\newblock {\em Physical Review E}, 69(6):066138, 2004.

\bibitem{Li2019}
Yuan Li, Pan Zhang, Yang Zhang, and Kunihiko Miyazaki.
\newblock Gait analysis using stereo camera in daily environment.
\newblock In {\em 41st Annual International Conference of the {IEEE}
  Engineering in Medicine and Biology Society, {EMBC} 2019, Berlin, Germany,
  July 23-27, 2019}, pages 1471--1475. {IEEE}, 2019.

\bibitem{Needham2021}
Laurie Needham, Murray Evans, Darren~P. Cosker, Logan Wade, Polly~M. McGuigan,
  James~L. Bilzon, and Steffi~L. Colyer.
\newblock The accuracy of several pose estimation methods for 3d joint centre
  localisation.
\newblock {\em Scientific Reports}, 11(1):20673, October 2021.

\bibitem{Mehdizadeh2021}
Sina Mehdizadeh, Hoda Nabavi, Andrea Sabo, Twinkle Arora, Andrea Iaboni, and
  Babak Taati.
\newblock Concurrent validity of human pose tracking in video for measuring
  gait parameters in older adults: a preliminary analysis with multiple
  trackers, viewing angles, and walking directions.
\newblock {\em Journal of NeuroEngineering and Rehabilitation}, 18(1):139,
  September 2021.

\bibitem{Muro-de-la-Herran2014}
Alvaro Muro-de-la Herran, Begonya Garcia-Zapirain, and Amaia Mendez-Zorrilla.
\newblock Gait analysis methods: An overview of wearable and non-wearable
  systems, highlighting clinical applications.
\newblock {\em Sensors}, 14(2):3362--3394, 2014.

\end{thebibliography}

\end{document}